\documentclass[11pt]{article}
\usepackage[letterpaper]{geometry}
\usepackage[parfill]{parskip}
\usepackage{amsmath,amsthm,amssymb,bbm}
\usepackage{mathtools}
\usepackage{cases}
\usepackage{graphicx}
\usepackage{caption}
\usepackage{subcaption}
\usepackage{tabularx}
\usepackage{microtype}
\usepackage{enumitem}
\usepackage{dsfont}
\usepackage[authoryear]{natbib}
\usepackage{algpseudocode,algorithm}

\usepackage{floatpag} 
\floatpagestyle{empty} 
\usepackage{authblk}
\usepackage{url}
\usepackage[colorlinks,citecolor=blue,urlcolor=blue,linkcolor=blue,linktocpage=true]{hyperref}
\usepackage{cleveref}
\crefformat{equation}{(#2#1#3)}
\crefrangeformat{equation}{(#3#1#4) to~(#5#2#6)}
\crefname{equation}{}{}
\Crefname{equation}{}{}

\makeatletter
\def\thm@space@setup{%
  \thm@preskip=\parskip \thm@postskip=0pt
}
\makeatother

\newtheorem{theorem}{Theorem}[section]
\newtheorem{lemma}[theorem]{Lemma}

\newtheorem{proposition}[theorem]{Proposition}
\newtheorem{definition}{Definition}
\newtheorem*{definition*}{Definition}
\newtheorem*{remark*}{Remark}
\crefname{definition}{\textbf{definition}}{definitions}
\Crefname{definition}{Definition}{Definitions}
\crefname{assumption}{\textbf{assumption}}{assumptions}
\Crefname{assumption}{Assumption}{Assumptions}

\newcommand{\pr}{\mathbb{P}}
\newcommand{\calM}{{\mathcal{M}}}

\let\hat\widehat
\let\tilde\widetilde

\begin{document}

\title{Differentially Private Model Selection \\ With
  Penalized and Constrained Likelihood}
\author[1]{Jing Lei\thanks{jinglei@andrew.cmu.edu}}
\author[2]{Anne-Sophie Charest}  
\author[3]{Aleksandra Slavkovic} 
\author[4]{Adam Smith} 
\author[1]{Stephen Fienberg}
\affil[1]{Department of Statistics, Carnegie Mellon University}
\affil[2]{D\'epartement de math\'ematiques et statistique,
    Universit\'e Laval}
\affil[3]{Department of Statistics, Pennsylvania
  State University}
  \affil[4]{Department of Computer Science and Engineering, Pennsylvania
  State University}

\maketitle 

\begin{abstract}

In statistical disclosure control, the goal of data analysis is twofold: The released information must provide accurate and useful statistics about the underlying population of interest, while minimizing the potential for an individual record to be identified. In recent years, the notion of \emph{differential privacy} has received much attention in theoretical computer science, machine learning, and statistics. It provides a rigorous and strong notion of protection for individuals' sensitive information. A fundamental question is how to incorporate differential privacy into traditional statistical inference procedures.
In this paper we study model selection in multivariate linear regression
under the constraint of differential privacy.  
We show that model selection procedures based on penalized least squares or likelihood can be made differentially private by a combination of regularization and randomization, and propose two algorithms to do so. We show that our private procedures are consistent under essentially the same conditions as the corresponding non-private procedures. We also find that under differential privacy, the procedure becomes more sensitive to the tuning parameters.   We illustrate and evaluate our method using simulation studies and two real data examples.
\end{abstract}



\section{Introduction}

In data privacy research, the goal of data analysis is to provide accurate and useful statistical inference while preventing individual records being identified. Such privacy protection is crucial in many statistical applications, namely for the analysis of census and survey data, medical and clinical studies, genetics data, and web user data collected on the internet.  In statistics, the treatment of confidential data has a long history under the name of ``Statistical Disclosure Control'' or ``Statistical Disclosure Limitation''; see for example \cite{Dalenius77}, \cite{Rubin93}, \cite{WillenborgD96}, \cite{fienberg10} and \cite{HundepoolD12}.  A long lasting challenge is to rigorously quantify the disclosure protection offered by privacy-preserving data analysis procedures.

The notion of \emph{differential privacy} has been introduced with the same objective in theoretical computer science by \cite{Dwork06} and \cite{DworkMNS06}. The general idea of differential privacy is to require for the outcome of a randomized data analysis procedure not to change much for small perturbations of the input data, so that one can not infer from the output the presence or absence of any individual in the input data set, or infer some of the person's characteristics. This requirement can be rigorously quantified, and does not depend on assumptions regarding the resources of the intruder, including any access to auxiliary information. Thus, differential privacy compares very favorably to measures of disclosure risk commonly used in the statistical disclosure control literature as it is more encompassing and a {\it worst case definition}, but at the same time it has been criticized as too stringent from the perspective of achieving needed statistical data utility; e.g., see \citet{FienbergRinYang, karwa2012psd}. 

Over the last decade, there has been a rapid development of differentially private algorithms and procedures in both the computer science and statistics literature.
For example, the main focus in the computer science literature was on designing differentially private mechanisms and efficient algorithms for private data release, e.g., the Laplace noise perturbation mechanism
 \citep{DworkMNS06}, the exponential mechanism \citep{McSherryT07}, releasing contingency tables \citep{BarakCDKMT07}, and boosting \cite{HardtLM10}. On the statistics side, research efforts include designing consistent and efficient differentially private point estimators \citep{DworkL09,Smith11,ChaudhuriMS11,Lei11,BassilyST14,KS16}, non-parametric density estimation \citep{WassermanZ10}, hypothesis testing \citep{FienbergSU11, JS13,USF13,YFSU14, KaSlKrPSD14, solea14, DSZ15,Shef15,WLK15,GLRV16}, and statistical lower bounds \citep{ChaudhuriH11,Duchi13}; there is also a large literature on private PAC learning, which echoes the concerns of statistical estimation in the context of classification~\citep{KLNRS08,BeimelKN10,BeimelBKN14,BeimelNS15,BunNSV15,Karwa2015private}.

In this paper we consider statistical model selection under the constraint of differential privacy. Despite the fast development in combining statistical theory and methodology with differential privacy, model selection under privacy constraints has not been well understood. 
The problem of differentially private model selection is motivated by practical concerns:
when private data analysis procedures are needed, it is rarely known which model is most appropriate for the data. When releasing the whole dataset is impractical due to privacy concerns and releasing point estimates for pre-specified models has limited utility, an appropriate compromise is to first identify the best model and then obtain and release consistent point estimates for this model. Both tasks need to be performed under privacy constraints, and thus we need methods for differentially private model selection. 

In particular, we focus on the classical linear regression model selection. In particular, we aim to provide insights for the following two questions.  (i) Is it theoretically possible to do model selection with differential privacy under the classical conditions? (ii) What new practical concerns arise in model selection when differential privacy is required?

We first show that the answer to the first question is positive by proposing differentially private model selection procedures based on penalized least squares and likelihood which exhibit asymptotic utility guarantee.  Here utility means that the procedure selects the correct model with high probability under appropriate regularity conditions.  In other words, differentially private model selection is theoretically possible in the classical setting.  More specifically, we propose a two-step differentially private model selection procedure: we first obtain a least square or maximum likelihood estimate under an $\ell_1$ constraint, then use noisy optimization with regularization to get the best model.  
Interestingly, the $\ell_1$ constraint, usually employed in high dimensional problems,
helps achieve privacy even in low dimensions.

Second, from simulations and real data examples we observe that the finite sample behavior of the proposed method depends crucially on the turning parameter required by the procedure.  For example, our algorithm imposes an $\ell_1$ norm constraint on the estimated regression coefficient. When the tuning parameter is conservatively chosen, i.e., the imposed upper bound of the $\ell_1$ norm is large, the utility is rather limited for the sample or moderate sample sizes.  When the sample size is in the thousands, the dependence of utility on tuning parameter is less significant, as predicted by the theory. More importantly, if auxiliary information is available, such as a good upper bound on the $\ell_1$ norm of the true regression coefficient, then we may be able to choose the tuning parameter more adaptively, with improved utility.  This reveals a distinct feature of differentially private data analysis: Some auxiliary information that does not affect classical inference may lead to significant performance improvement in differentially private analysis. Indeed, the $\ell_1$ bound on the regression coefficient which improves the utility of our differentially-private procedure is useful in Lasso when the dimensionality is high, but not so much in the classical regime. 

The remainder of the paper is organized as follows: In section 2, we review the definition and interpretation of differential privacy, as well as two general methods to create differentially private algorithms. Section 3 details the two proposed differentially private model selection procedures, with proofs of their privacy guarantee, and notes on the choice of the tuning parameters. Statements and proofs of the utility guarantee of the two algorithms are given in section 4.  Empirical results, including a simulation study and two real data examples are reported in section 5. Section 6 provides a brief discussion.  All proofs and technical details are collected in section 7.

\section{Differential privacy}

Differential privacy requires that the output of a procedure is not drastically altered under small perturbations of the input data set, such that an attacker, regardless of his auxiliary information and
computing power, can hardly recover the presence/absence of a particular individual in the 
data set.  The notion of differential privacy is a property of that data analysis procedure, rather than of the output obtained. 

\subsection{Definition}

To formalize, consider a data set $D=\{z_1,...,z_n\}\in
\mathcal Z^n$ consisting of $n$ data points in sample space $\mathcal Z$. A
 data analysis procedure $\mathcal T$, possibly randomized, maps the data set $D$, together with a random input $\omega$, to $\mathcal T(D)\equiv\mathcal T(D,\omega)\in S$, an output space. Here we assume that $(S,\mathcal S)$ is  a measurable space and
$\mathcal T(D,\cdot):\Omega\mapsto S$ is a measurable function.
 Whenever it is not confusing, we will use $\mathcal T(D)$ to
denote the random variable $\mathcal T(D,\omega)$.

For any two data sets $D$ and $D'$  of the same size, we use
${\rm Ham}(D,D')$ to denote their \emph{Hamming distance}, the
number of entries at which they differ, regardless of the order.
For example, if $\mathcal Z = $ $\mathbb R$ and $D=\{1,2,3,4\}, D'=\{2,3,4,5\}$, then ${\rm Ham}(D,D')=1$. In the rest of this paper, we always use $D$ and $D'$ to denote a pair of adjacent
data sets that differ at the last entry: $D=\{z_1,...,z_{n-1},z_n\}$, $D'=\{
z_1,...,z_{n-1}, z_n'\}$.

We can now state formally the property of differential privacy: 
\begin{definition}[$\epsilon$-differential privacy, \cite{DworkMNS06}]
  \label{def:dp}
  Given a privacy parameter, $\epsilon>0$, the procedure $\mathcal T$ satisfies \emph{$\epsilon$-differential privacy}
  if
  $$\sup_{{\rm Ham}(D,D')=1,A\in\mathcal S} 
   \left|\log\frac{P_{\omega}(\mathcal T(D',\omega)\in A)}
  {P_\omega(\mathcal T(D,\omega)\in A)}\right|
  \le \epsilon\,$$
  where we define $\log \frac{0}{0}=0$ for convenience.
\end{definition}

  In the above notation, $P_\omega$ denotes the probability with respect to
   $\omega$, which is the source of randomness in the data analysis
   procedure. Thus, the definition does not impose any conditions on
   the distribution of $D$ --- the privacy is required to hold for \emph{all
   pairs of adjacent data sets}. This stringent condition means that
   procedures which satisfy definition~\ref{def:dp} provide very strong privacy
   guarantees, even against adversaries who have considerable partial
   information about the data set \citep{KasiviswanathanS08}. 
In order to satisfy this definition, any
   nonconstant procedure must be randomized. 

The $\epsilon$-differential privacy is a strong requirement, as it takes
supremum over all possible neighboring data sets of size $n$.
A mild relaxation is the $(\epsilon,\delta)$-differential privacy.
\begin{definition}[$(\epsilon,\delta)$-differential privacy]
  Given $\epsilon>0$, $\delta\in(0,1)$, a procedure satisfies
  $(\epsilon,\delta)$-differential privacy if, for all 
  measurable $A\subseteq \mathcal S$ and all neighboring data sets $D$, $D'$,
  $$
  P_{\omega}(\mathcal T(D)\in A)\le e^{\epsilon} P_{\omega}(\mathcal T(D)\in A)+\delta\,.
  $$
\end{definition}
Here the requirement of the original $\epsilon$-differential privacy is relaxed
so that the distribution of $\mathcal T(D)$ only needs to be dominated by that
of $\mathcal T(D')$ outside of a set with probability no more than $\delta$.

 In our discussion of statistical applications we
   will generally focus on data sets consisting of a sequence of $n$
   independent random samples from an underlying distribution, and the
   corresponding probability will be denoted $P_D$.  Note that differential privacy definition has no such assumption. We will
   use $P_{D,\omega}$
   to denote the overall randomness due to both the data and random
   mechanism in the analysis.



\subsection{Statistical interpretation of differential privacy}

Privacy protection due to differential privacy can be interpreted from
a Bayesian perspective, as in \cite{AbowdV08,kasi08bayes}. Suppose one has a prior
distribution of the input data set $D$ and then gets to observe the
random output $\mathcal T(D,\omega)$.  Denote by $P_i$ and $Q_i$ the marginal prior and posterior of $X_i$, for the $i$th entry in $D$. Then if the procedure $\mathcal T$ is differentially-private and the prior $P$ is a product measure on the entries of $D$, we have that $e^{-\epsilon}\le d P_i/d Q_i\le e^\epsilon $, thus limiting the information regarding $X_i$ gained from the output $\mathcal T(D,\omega)$.  A related hypothesis testing interpretation is given in Theorem 2.4 of \cite{WassermanZ10}.

One may also intuitively interpret differential privacy as a specific notion of
robustness.  It requires that the distribution of $\mathcal T(D)$ is not changed too much if $D$ is perturbed in only one entry. However, the definition of differential privacy is also a \emph{worst case definition}, where the probability ratio needs to be uniformly bounded over all pairs of adjacent input data sets.  
A key ingredient to designing differentially private statistical procedures
is to bridge the gap between worst case privacy guarantee and the average
case statistical utility.  It turns out that robustness and regularization
are the most relevant structures to explore, as in \cite{DworkL09}, \cite{ChaudhuriMS11} and 
\cite{ST13}.

\subsection{Designing differentially private algorithms}\label{sec:generic}

For any statistical task which we want to carry, a specific randomized procedure $\mathcal T$ must be designed to take as input a database $D \in \mathcal Z^n$ and return an element of the output space $\mathcal S $ while satisfying differential privacy. There exists a few approaches which are generic enough to be adaptable to various tasks, and which are often used as building blocks for more complicated procedures. We present two of these methods, which we use later in the construction of our procedure. The first adds random noise to a non-private output and the second samples randomly from a set of candidate outputs.

\paragraph{Adding Laplace noise}  The additive noise approach is
applicable when the output space is an Euclidean space.
For presentation simplicity we consider here $S=\mathbb R$. 
Let $T(D)$ be a non-private
mechanism. Define the \emph{global sensitivity} $G_T$ as
\begin{equation}\label{eq:global-sensitivity}
  G_T=\sup_{{\rm Ham}(D,D')=1} |T(D)-T(D')|\,.
\end{equation}
If $G_T<\infty$, then it is easy to check \citep{DworkMNS06} that
 \begin{equation}\label{eq:add-laplace-gs}
 \mathcal T(D)\equiv T(D)+ \epsilon^{-1}G_T\zeta 
 \end{equation} satisfies
$\epsilon$-differential privacy, where $\zeta$ is a standard double exponential random variable
with density function $0.5 \exp(-|\zeta|)$ (also known as the Laplace distribution).

%

\paragraph{The exponential mechanism and noisy optimization} The \emph{exponential mechanism}
\citep{McSherryT07}
is designed for discrete output spaces.  Suppose $S=\{s_\alpha:\alpha\in\aleph\}$ and
let $q:\mathcal Z^n \times S \mapsto \mathbb R$ be a score function that measures
the quality of $s\in S$ in terms of its agreement with the input data set.  
Usually $q(D, s)=-|s-T(D)|$
for some deterministic procedure $T$.  Denote $G_q=\sup_s G_{q(\cdot,s)}$, 
where $G_{q(\cdot,s)}$ is the global sensitivity of the mapping $q(\cdot,s)$.
Let $\mathcal T(D)$ be the procedure that outputs a random sample from $S$ with 
probability
$$P_\omega(\mathcal T(D)=s)\propto \exp\left(\frac{\epsilon q(D,s)}{2G_q}\right)\,.$$
Then $\mathcal T$  satisfies $\epsilon$-differential privacy.

When $S$ is finite, one may also use additive noise to approximately maximize $q(D,s)$ over 
$s$.
Let $$\tilde q(D,s)=q(D,s)+2\epsilon^{-1}\zeta G_{q(\cdot,s)}$$ be the privatized
score, where $\zeta$ is an independent draw from the standard double exponential distribution.
Then $$\mathcal T(D)=\arg\max_s \tilde q(D,s)$$ satisfies $\epsilon$-differential privacy and 
usually offers similar performance as the exponential mechanism. 

\paragraph{A generic scheme for $(\epsilon,\delta)$-differential privacy}
In many statistical problems the sample space is not compact and hence $G_T=\infty$ for
many statistics $T$ such as the sample mean.  A general strategy is to show that
one can add much less noise for most average case data sets with ($\epsilon$, 
$\delta$)-differential privacy \citep{DworkL09}.  These methods often involves the
notion of \emph{local sensitivity} of a deterministic procedure $T$:
\begin{equation}
  G_{T}(D) = \sup_{D': {\rm Ham}(D,D')=1} |T(D)-T(D')|\,.
\end{equation}
If $G_T(D)$ is finite and public, then one can show that
adding noise to $T(D)$ as in \eqref{eq:add-laplace-gs} with $G_T$ replaced by
$G_T(D)$ also gives $\epsilon$-differential privacy.  Unfortunately, $G_T(D)$
depends on the data set and may contain sensitive information.
However, there is a generic scheme based on this idea with valid privacy guarantee
under the following two general conditions on the deterministic procedure $T$.
 \begin{enumerate}
   \item For all $\epsilon>0$ there exists a real-valued function $G^*(D)$ and a 
   randomized
   procedure $\mathcal{T}_{\epsilon}(D,g)$,
    which is $\epsilon$-differentially private
   if $g\ge G^*(D)$ and is assumed to be non-private.
   \item Given $\epsilon>0$ and $\delta\in(0,1)$,
   there exists an $\epsilon$-differentially private mapping
   $G_{\epsilon,\delta}(D)$ satisfying
   $P_\omega[G_{\epsilon,\delta}(D)\ge G^*(D)]\ge 1-\delta$ for all $D$.
 \end{enumerate}
 An example of $G^*(D)$ is the local sensitivity of some procedure $T(D)$, and $\mathcal T$
 is the noisy version as in \eqref{eq:add-laplace-gs} calibrated to a upper bound of the 
 local sensitivity.
 
\begin{proposition}\label{pro:generic}
  Under the above two assumptions, 
  for any $\epsilon_1+\epsilon_2=\epsilon$, and $\delta\in (0,1)$,
   $\mathcal T_{\epsilon_2}(D,G_{\epsilon_1,\delta}(D))$
   satisfies $(\epsilon,\delta)$-differential privacy.
\end{proposition}

\section{Differentially Private Model Selection Procedures}
\label{sec:methodology}

Popular methods of model selection for linear regression include information
criteria such as AIC \citep{AIC} and BIC \citep{BIC}, cross-validation \citep{Picard84}, and the more recent penalized least squares \citep[e.g.,][]{Lasso,SCAD}.
In this paper we focus on penalized least squares and penalized profile likelihood estimators, both being variants of the classical approach based on information criteria.

 \subsection{Background: Linear regression and model selection}
 \label{background}
 
In the linear regression model, the data points are independent, each consisting of a 
response $Y\in \mathbb R^1$,
and a covariate $X\in\mathbb R^d$, which satisfies
\begin{equation}\label{eq:linear-model}
Y=X^T\beta_0 + W
\end{equation}
where $\beta_0\in\mathbb R^d$ is the regression coefficient, and $W$ is a Gaussian random 
variable, independent of $X$, with mean zero and variance $\sigma^2$.
The observed data set is $D=\{(X_i,Y_i):1\le i \le n\}$, where $X_i=(X_{i1},...,X_{id})^T$.

The model selection problem is to find the support of 
$\beta_0$: $M_0\equiv \{j:\beta_0(j)\neq 0\}$.
 To give a precise formulation, consider a class of candidate models
$\mathcal M\subseteq \{0,1\}^{d}$ that contains $M_0$.  For each $M\in \mathcal M$,
the corresponding hypothesis is $\beta_0\in\Theta_M\equiv \{\beta\in\mathbb R^d:
\beta_0(j)=0,~\forall~j\notin M\}$.

Given a parameter $(\beta,\sigma^2)$ and an observed data set, the log likelihood
is, ignoring constant terms,
$$
\ell(\beta,\sigma^2;D)=\sum_{i=1}^n \left[ -\frac{1}{2}\log \sigma^2 
-\frac{1}{2\sigma^2}(Y_i-X_i^T\beta)^2 \right]\,.
$$
We consider two cases separately.

\paragraph{Known variance.}
When $\sigma^2$ is known, we aim to find the $\beta$ that maximizes the log likelihood, which leads to model selection with the least squares. Without loss of generality, we assume $\sigma^2=1$. We then define
\begin{align}
\ell(\beta;D) = & -\frac{1}{2}\sum_{i=1}^n (Y_i-X_i^T\beta)^2\,,\\
\hat\beta_M = & \arg\min_{\beta\in\Theta_M} -2\ell(\beta;D)\,,\\
\ell(M;D) = &    \ell(\hat\beta_M;D)\,.
\end{align}



\paragraph{Unknown variance.}
In the more realistic setting that $\sigma^2$ is unknown, we can maximize over the nuisance parameter $\sigma^2$ to perform model selection with the profile likelihood. Ignoring constant terms, we obtain the profile log-likelihood for $\beta$:
\begin{equation}
\ell^*(\beta;D)=-\frac{n}{2}\log
\left[\frac{1}{n}\sum_{i=1}^n(Y_i-X_i^T\beta)^2\right]\,,
\end{equation}
Maximizing this over all $\beta$ in $\Theta_M$, we get
\begin{equation}
\ell^*(M;D)\equiv \sup_{\beta\in\Theta_M}\ell^*(\beta;D).
\end{equation}

\paragraph{Information criteria.}
One could simply maximize $\ell(M;D)$ or $\ell^*(M;D)$ as given above. However, more complex models tend to  yield larger values of log likelihood. The approach of information criteria thus minimizes the sum of negative log likelihood and a measure of model complexity. For example, in the case of unknown $\sigma^2$, one choose the model by
\begin{equation}\label{eq:information-criteria}
\hat M = \arg\min_{M\in \mathcal M} -2\ell^*(M;D) + \phi_n |M|\,,
\end{equation}
where $\phi_n$ is the amount of penalty on the model complexity. The best known and most widely used examples are the AIC ($\phi_n=2$), and BIC ($\phi_n=\log n$).

In the case of known $\sigma^2$, the penalized minimization becomes
\begin{equation}\label{eq:penalized-least-square}
\hat M = \arg\min_{M\in \mathcal M} -2\ell(M;D) + \phi_n |M|\,. 
\end{equation}
It is worth noting that the penalized least square estimator, with appropriate choice of $\phi_n$, can also be used in the case of unknown $\sigma^2$.


\subsection{Towards Differentially Private Model Selection}
To perform model selection in a differentially-private manner, we propose to apply the exponential mechanism or noisy minimization to the minimization problems \eqref{eq:information-criteria} and \eqref{eq:penalized-least-square},  with $q(D,M)$ (here $s$ in the general definition is replaced by $M$) given by the penalized likelihood 
$$L^*(M;D)\equiv -2\ell^*(M;D)+\phi_n|M|\,$$
in the case of penalized log profile likelihood \eqref{eq:information-criteria},
or
$$
L(M;D)\equiv -2\ell(M;D)+\phi_n|M|
$$
in the case of penalized least squares \eqref{eq:penalized-least-square}.

A key step in this approach is to evaluate and control the sensitivities of $L(M;D)$ and $L^*(M;D)$
as functions of $D$ for any $M\in\mathcal M$. 

To simplify the notation and concentrate on the main idea, we will assume in the sequel that the data entries are bounded or standardized:
\begin{equation}
	\label{eq:bound_on_data}
	\max_{1\le i\le n}|Y_i|\le r,\quad \max_{1\le i\le n, 1\le j\le d}|X_{ij}|\le 1\,,
	\end{equation}
where $r$ is a known number that can grow with $n$. Boundedness is typically required for differentially private data analysis.  Standard methods for finding the range of the data set in a privacy-preserving manner  include those given in \cite{DworkL09} and \cite{Smith11}.

\subsubsection{Sensitivity of least squares and profile log-likelihood}

As indicated in Section \ref{background}, we consider procedures based on two score functions for a model: the sum of squared residuals  $\ell(M;D)$, and the profile
likelihood $\ell^*(M;D)$. 

To bound the sensitivity of either of them, we must bound the possible parameter vectors we consider. We thus consider constrained versions of the two score functions. Given $R>0$, we define 
\begin{equation}
-2\ell_{R}(M;D)=\min_{\beta\in\Theta_{M},\|\beta\|_1\le R}-2\ell(\beta,D)\,.\label{eq:1}
\end{equation}
and 
\begin{equation}
-2\ell^*_{R}(M;D)=\min_{\beta\in\Theta_M,\|\beta\|_1\le R}
-2\ell^*(\beta;D)\,.
\end{equation}


The sensitivity of the least squares loss  is now easy to
bound: 

\begin{lemma}[Sensitivity of constrained least-squares]\label{lem:sensitivity-LSE}
  When $\beta$ has $\ell_1$-norm at most $R$ and the data $(X_i,Y_i)$
  are restricted to
  $[-1,1]^d\times [-r,r]$, the global sensitivity of the squared error loss functions
  $-2\ell(\beta;\cdot)$ and  $-2\ell_R(M;\cdot)$ is 
  at most $(r+ R)^2$.
\end{lemma}
The proof is both short and elementary, and hence omitted.

In the case of the constrained profile likelihood, we can not bound the global sensitivity, but we can find a bound on the local sensitivity:

\begin{lemma}[Local sensitivity of constrained the profile likelihood]\label{lem:local-sensitivity-RML}
 Under the same conditions as in Lemma \ref{lem:sensitivity-LSE}, the local sensitivity of $-2\ell^*_R(M;\cdot)$ is 
  no larger than
  $$
  \frac{n(r+R)^2}{-2\ell_R(M;D)-(r+R)^2}\,.
  $$
\end{lemma}

\subsubsection{Algorithms} 
Lemma \ref{lem:sensitivity-LSE} implies that the penalized constrained lease squares
$$L_{R}(M;D)=-2\ell_{R}(M;D)+|M|\phi_n,$$
can easily be minimized in an $\epsilon$-differentially private manner. The complete algorithm is given below.

\begin{algorithm}[tb]
    \caption{Model Selection via Penalized Constrained Least Squares (PCLS)}
    \label{alg:penalizedSSR}
\begin{algorithmic}
  \State{\bf Input:} Data set $D=\{(X_i,Y_i): i \in \{1,...,n\}\}$, collection of
    models $\calM$, parameters
    $r, R, \phi_n, \epsilon$.

  \State{\bf Output:} Estimated model $\hat M \in \calM$

  \For{each model $M$ in $\calM$}
    \State $\displaystyle \ell_R(M;D) \gets \max_{\beta\in\Theta_M,
      \|\beta\|_1\leq R}
    -\frac{1}{2}\sum_{i=1}^n (Y_i-X_i^T\beta)^2$\;
     \State $\displaystyle L_R(M;D) \gets -2\ell_R(M;D) + \phi_n |M| $ \;  
    \State $\tilde L_R(M;D) \gets L_R(M;D) + \frac{2(r+R)^2}{\epsilon} Z_M,\;$  where $Z_M$ is a Laplace random variable\;
\EndFor
  \State Return $\displaystyle\arg\min_{M \in \calM} \tilde L_R(M;D)$
\end{algorithmic}
\end{algorithm}


Next, Lemma \ref{lem:local-sensitivity-RML} gives an upper bound of the local sensitivity of $L_R^*(M;D)$.  Now we apply the generic scheme of designing $(\epsilon,\delta)$-differentially private algorithms described in section \ref{sec:generic}.

First let 
  $$
  G^*(D) \equiv \frac{n(r+R)^2}
  {\min_{M}-2\ell_{R}(M;D)-(r+R)^2}\,.
  $$
Then $G^*(D)$ is an uniform upper bound of the local sensitivity of $L_R(M;D)$ for all $M$.
The only private part in $G^*(D)$ above is $\min_{M\in\mathcal M}-2\ell_{R}(M;D)$,
which has global sensitivity $(r+R)^2$ according to Lemma \ref{lem:sensitivity-LSE}. Thus following the general procedure in Section 2.3, a valid choice of $G(D)$ for $-2\ell^*_{R}(M;D)$
is
$$
G(D)=\frac{n(r+R)^2}
{\min_{M}-2\ell_{R}(M;D)-(r+R)^2+
\epsilon^{-1}(r+R)^2\left(Z_G-\log\frac{1}{2\delta}\right)}
$$
where $Z_G$ is a standard Laplace random variable.

Then we can construct the final estimators using either exponential mechanism or the noisy minimization, both satisfying $(2\epsilon,\delta)$-differential privacy.
The complete algorithm is given in Algorithm 2.

\begin{algorithm}[tb]
    \caption{Model Selection via Penalized Constrained Profile Likelihood (PCPL)}
    \label{alg:profile}
    \begin{algorithmic}
  \State{\bf Input:} Data set $D=\{(X_i,Y_i): i \in \{1,...,n\}\}$, collection of
    models $\calM$, parameters
    $r, R, \phi_n, \delta, \epsilon$.

  \State{\bf Output:} Estimated model $\hat M \in \calM$

\For{each model $M$ in $\calM$}
  
  \State  $\displaystyle \ell^*_R(M;D) \gets \max_{\beta\in\Theta_M,
      \|\beta\|_1\leq R}
    -\frac{n}{2} \log \left[ \frac{1}{n} \sum_{i=1}^n (Y_i-X_i^T\beta)^2 \right] $\;
    
  \State  $L^*_R(M;D) \gets \ell^*_R(M;D) + \phi_n |M| $ \;  

  \State    $ \tilde L^*_R(M;D)\gets L^*_R(M;D) + \frac{2G(D)}{\epsilon} Z_M $  where
      $$  G(D)=\frac{n(r+R)^2} {\min_{M}-2\ell_{R}(M;D)-(r+R)^2+ \epsilon^{-1}(r+R)^2\left(Z_G-\log\frac{1}{2\delta}\right)} ; $$ 
\hspace{0.35 in} 
$\ell_{R}(M;D)$ is as in Algorithm 1, and $Z_G$, $Z_M$ are Laplace random variables;
  \EndFor
  \State Return $\arg\min_{M \in \calM} \tilde L^*_R(M;D)$      
    \end{algorithmic}
\end{algorithm}


\subsection{Choosing the tuning parameters}
Both algorithms introduced above require two tuning parameters:
an upper bound $R$ of the $\ell_1$ norm of the regression coefficient, and the penalty parameter $\phi_n$.
In the context of privacy-preserving data analysis, there are two requirements on the quality of any inference procedure: privacy and utility.
Both proposed methods satisfy their corresponding $\epsilon$-differential privacy (Algorithm 1) and $(2\epsilon,\delta)$-differential privacy (Algorithm 2) for any choices of $R$ and $\phi_n$.
Regarding utility, our theoretical analysis shows that both algorithms achieve consistent model selection for a wide range of $R$ and $\phi_n$.

However, our numerical experiments in section 5 show that the performance of our proposed algorithms is sensitive to the choice of these tuning parameters. We acknowledge that fully data-driven and privacy-preserving methods for choosing these tuning parameters remain a challenging and important open problem, and is beyond the scope of this paper. Data-driven choice of penalty parameter is a hard problem even without privacy constraint. Here we provide some heuristics on potential solutions.

Regarding choosing $R$, the ideal choice would be the $\ell_1$ norm of the true regression coefficient.
In practice, a good choice of $R$ should be close to $\|\beta_0\|_1$.
This can be achieved by finding a differentially private estimate of the maximum $\ell_1$ norm of $\hat \beta_M$ over all $M\in \mathcal M$, which is feasible if only a single number is released.

Regarding choosing $\phi_n$, a good choice of $\phi_n$ needs to be large enough so that it dominates any
statistical sampling noise in the data and the additive noise due to privacy constraints.  On the other hand, $\phi_n$ cannot be too large, because otherwise it will introduce substantial bias in the selected model.  Taking the penalized least squares estimator for example, a good $\phi_n$ shall roughly be the difference of the largest $\ell(M;D)$ and the second largest $\ell(M;D)$, which can possibly be estimated with differential privacy as a single number.

  
%


\section{Utility Analysis}
In privacy-preserving data analysis, the privacy shall be protected for
\emph{any} possible input data set.  In other words, the privacy guarantee
needs to cover the worst case and must
 be established with no distributional conditions on the data set.
 In the previous sections, our differentially private procedure only
 requires the data to be bounded, which can be verified or enforced easily
 in practice.
 On the other hand, the statistical utility (for example, consistency, rate of
 convergence) is usually based on common statistical assumptions on the data.
To facilitate discussion, we first introduce some notation and assumptions.

\paragraph{Notation.} Let $\mathbf X$ be the $n\times d$ design matrix, and $\mathbf Y$ the $n\times 1$ vector
of $Y$'s.  For any $M\in \mathcal M$, let $\mathbf X_M$ be the $n\times |M|$ design matrix
consisting of the columns in $M$.  Then
 $\hat\beta_M$ is a $d\times 1$ vector that is
$(\mathbf X_M^T\mathbf X_M)^{-1}\mathbf X_M^T \mathbf Y$
on the entries in $M$ and zero elsewhere. It is
the ordinary least square estimate under model $M$.
The sample covariance is $\hat \Sigma=n^{-1}\mathbf X^T\mathbf X$.

\paragraph{Assumptions.}
We state below five assumptions required for our utility results, and discuss their practical significance. \\

Our first assumption is a sparse linear model with Gaussian noise.
\begin{itemize}
  \item [\textbf{A0}.] (Linear model with Gaussian noise) The data entries $(X_i,Y_i)_{i=1}^n$
 are generated from the linear regression model \eqref{eq:linear-model} with
 a regression coefficient vector $\beta_0$ with $d_0=\|\beta_0\|_0$ and $b_0\equiv \min_{j:\beta_0(j)\neq 0}|\beta_0(j)|$. The noise $W_i$ are
 iid Gaussian with mean zero and variance $\sigma^2$.
\end{itemize}



Next we assume that the number of candidate models grows polynomially with sample size $n$. This is usually the case when we only search over sparse models and the total number of variables grows polynomially in $n$.
\begin{itemize}
  \item [\textbf{A1}.] (Candidate models) The set of candidate models $\mathcal M$ contains the true model $M_0=\{j:\beta_0(j)\neq 0\}$, and has cardinality no more than $n^\alpha$ for some positive number $\alpha$ which is allowed to grow with $n$.
The largest candidate model has $\bar d$ variables.
\end{itemize}

The worst-case sensitivity of the log likelihood is hard to control because the
design matrix $\mathbf X$ may be poorly conditioned.  Thus we need to add
some singular value condition on the design matrix.
\begin{itemize}
  \item [\textbf{A2}.] (Design matrix) The design matrix $\mathbf X$ is fixed, and the sample covariance satisfies the sparse eigenvalue
  condition:
  $$\kappa_0\equiv\inf_{1\le\|\beta\|_0\le
  \bar d+d_0}\frac{\beta^T\hat\Sigma\beta}{\|\beta\|_2^2}>0.$$
\end{itemize}
  Assumption A2  excludes the situation of linear dependence between
  columns of $\mathbf X_{M_0}$ and $\mathbf X_{M_0^c}$.  A similar condition has been considered
  in the literature on model selection consistency using information criteria 
  \citep{Nishii84}.
  The condition stated here is for fixed design matrix $\mathbf X$, but it holds with high 
  probability for many random designs.
  Assumption A2 also implies an upper bound of the range of $\hat\beta_M$. Such a boundedness
property will help control the sensitivity of the log likelihood.

Moreover, as we did in section 3, we assume boundedness of the data entries, which are typically needed for developing differentially private procedures.
\begin{itemize}
  \item [\textbf{A3}.] (Boundedness) Each entry of $\mathbf Y$ is bounded by $r$, and each entry of $\mathbf X$ is bounded by $1$.
\end{itemize}

Finally, we assume that the sample size is large enough, when compared to other quantities in the analysis such as the noise variance, and the inverse of privacy parameter $\epsilon$.
\begin{itemize}
  \item [\textbf{A4}.] (Sample size) The sample size $n$ is large enough so that equations \eqref{eq:detail-assumption-1}, \eqref{eq:detail-assumption-2}, and \eqref{eq:detail-assumption-3} hold.
\end{itemize}

\subsection{Utility of Penalized Constrained Least Squares}
We first give  a utility result for the noisy penalized constrained least squares estimator (Algorithm 1).
%
%


\begin{theorem}\label{thm:utility-pcls}
  Assume A0-A4 hold. If $\phi_n$ satisfies
  $$
  2A(1\vee \sigma^2)\log n \le \phi_n \le \frac{1}{4\vee(1+2d_0)}\kappa_0 b_0^2\sigma^2 n\,,
  $$
  where $A=2(\alpha+c)$ for some $c>0$,
  and
  $$R\ge r\sqrt{\frac{\bar d}{\kappa_0}}\,,$$
  then the PCLS estimator $\hat M$ given by Algorithm \ref{alg:penalizedSSR} with privacy parameter $\epsilon$ and
  penalty parameter $\phi_n$ satisfies
  $$
  P_{D,\omega}(\hat M\neq M)\le 
  n^{\alpha}\exp\left(-\frac{\phi_n\epsilon}{4(R+r)^2}\right)+\frac{1+2\bar d}{\sqrt{2\pi A\log n}}n^{-c}\,.
  $$
\end{theorem}
Recall that  $P_{D,\omega}$ stands for taking probability over both the randomness in $D$ and in the generation of Laplace random variables (denoted by $\omega$) in the algorithm.  Here we are assuming a fixed design matrix, so the randomness in $D$ is equivalent to the randomness of the additive noise $W$.

\subsection{Utility of Penalized Constrained Profile Likelihood}
Now we provide utility result for the noisy penalized profile likelihood estimator.

\begin{theorem}
  [Utility of penalized profile likelihood]\label{thm:utility-local-sensitivity}
  Assume A0-A4 hold.
  If $\phi_n$ satisfies
   $$4A\log n \le \phi_n\le \frac{2}{3|M_0|}n\log\left(1+\frac{\kappa_0b_0^2}{4\sigma^2}\right)\,,
$$
where $A=2(\alpha+c)$ for some $c>0$,
and
$$R\ge r\sqrt{\frac{\bar d}{\kappa_0}}\,$$
then for any constant $c>0$ and $n$ large enough as quantified in \Cref{eq:detail-assumption-1,eq:detail-assumption-2,eq:detail-assumption-3}, the
selected model $\widehat M$ given by the noisy penalized constrained profile likelihood (Algorithm \ref{alg:profile}) satisfies
 $$P_{D,\omega}\left(\hat M\neq M_0\right)
 \le n^{\alpha}\exp\left(-\frac{\epsilon\sigma^2\phi_n}{64(R+r)^2}\right)+3n^{-A/2}+2\bar d n^{-c}\,.$$
\end{theorem}


 \section{Empirical Results}

We have shown in the last section that the two proposed differentially-private model selection procedures are consistent under similar conditions as the corresponding non-private algorithms. We now provide some empirical results for model selection via penalized constrained least squares, to illustrate both the utility of the algorithms and impact of tuning parameter selection as discussed in section 3.3.

 \subsection{Simulation study}

We consider two different regression models, both of the form $Y=X^T\beta_0 + W,$ where $X$ is an $n$ by 6 matrix with columns sampled independently from the uniform distribution on $[-1,1]$ and $W$ is standard normal, but the regression coefficients are different. The first model uses $\beta_0 = (1,1,1,0,0,0)$ whereas the second model uses $\beta_0 = c(1.5,1,0.5,0,0,0)$. Note that the $\ell_1$ norm of $\beta_0$ is $3$ in both cases, and thus the oracle value for $R$. However, the first model should be easier to recover from data. 

In the simulations, we consider a small sample size ($n=100$) and a moderately large sample size ($n=1000$), as well as $R \in \{1,2.5,3.5,10\}$, $\epsilon \in \{0.1, 1, 5, 10\}$ and $\phi \in [0,n/2]$. For each set of parameters, we sample $500$ data sets from the true model, apply Algorithm 1 to each of them, and note the proportion of times we correctly identify the correct model. We set $r$ to be the maximum observed value of $Y$ in the data set. 

Figure 1 shows the results for model 1 where $\beta_0 = (1,1,1,0,0,0)$. Except for with the smallest value of $R$ the procedure is very successful at choosing the correct model over a wide range of $\phi$ values. The poor results for $R=1$ are explained by the fact that since $R$ is much smaller than the true $\ell_1$ norm of $\beta_0$, the parameter can not be estimated properly. When $R$ is too large, the estimation of $\beta_0$ is unchanged, but a larger amount of noise is needed to satisfy differential privacy, which explains the small drop in utility. As expected, decreasing $\epsilon$ needed to provide stricter privacy also leads to decreased utility. Note however that for $n=1000$ the method is very accurate for a wide range of $\phi$ even for $\epsilon$ as small as $1$. The results also confirm our claim that the choice of $\phi$ is less crucial for larger sample size, as the procedure works well on an entire interval. Note also that the actual value of $\phi$ depends on $n$, with the optimal $\phi$ increasing roughly linearly with $n$. 

Figure 2 shows the results for case where $\beta_0 = c(1.5,1,0.5,0,0,0)$. The effect of $n$, $R$, $\phi$ and $\epsilon$ are very similar in this case, but the choice of $R$ and $\phi$ are more important to achieve good utility. The sensitivity of the procedure to choices of $R$ and $\phi$ thus depends on the structure of the true parameter $\beta_0$. Note however that for $n=1000$ a proper choice for the parameters leads to completely accurate model selection with $\epsilon = 5$, and even very accurate for $\epsilon=1$ with $R=2.5$. 

\begin{figure}[t]
\centering
\begin{subfigure}{.55\textwidth}
  \centering
  \includegraphics[scale = 0.4]{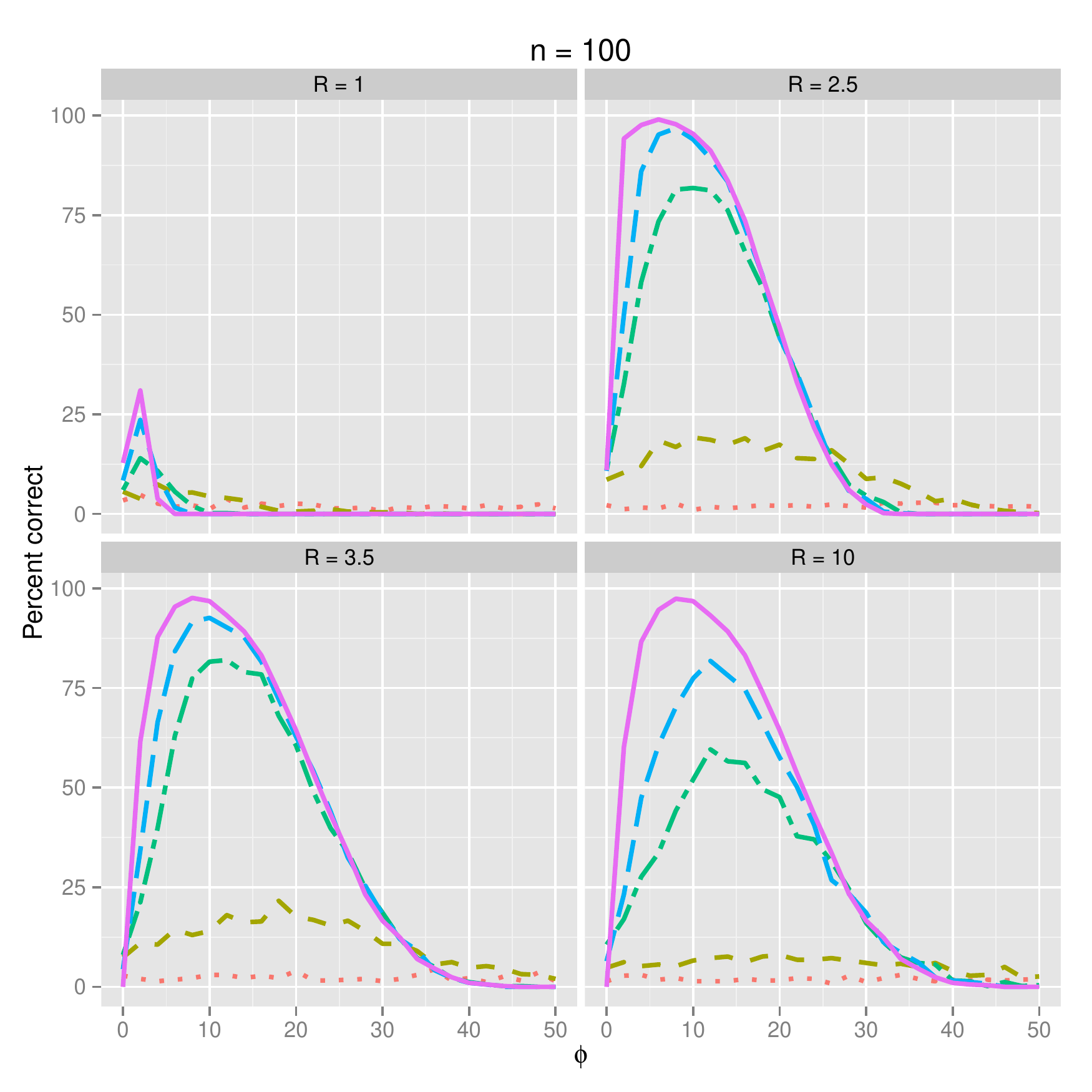}
  \caption{}
  \label{fig:sub1}
\end{subfigure}%
\begin{subfigure}{.55\textwidth}
  \centering
  \includegraphics[scale = 0.4]{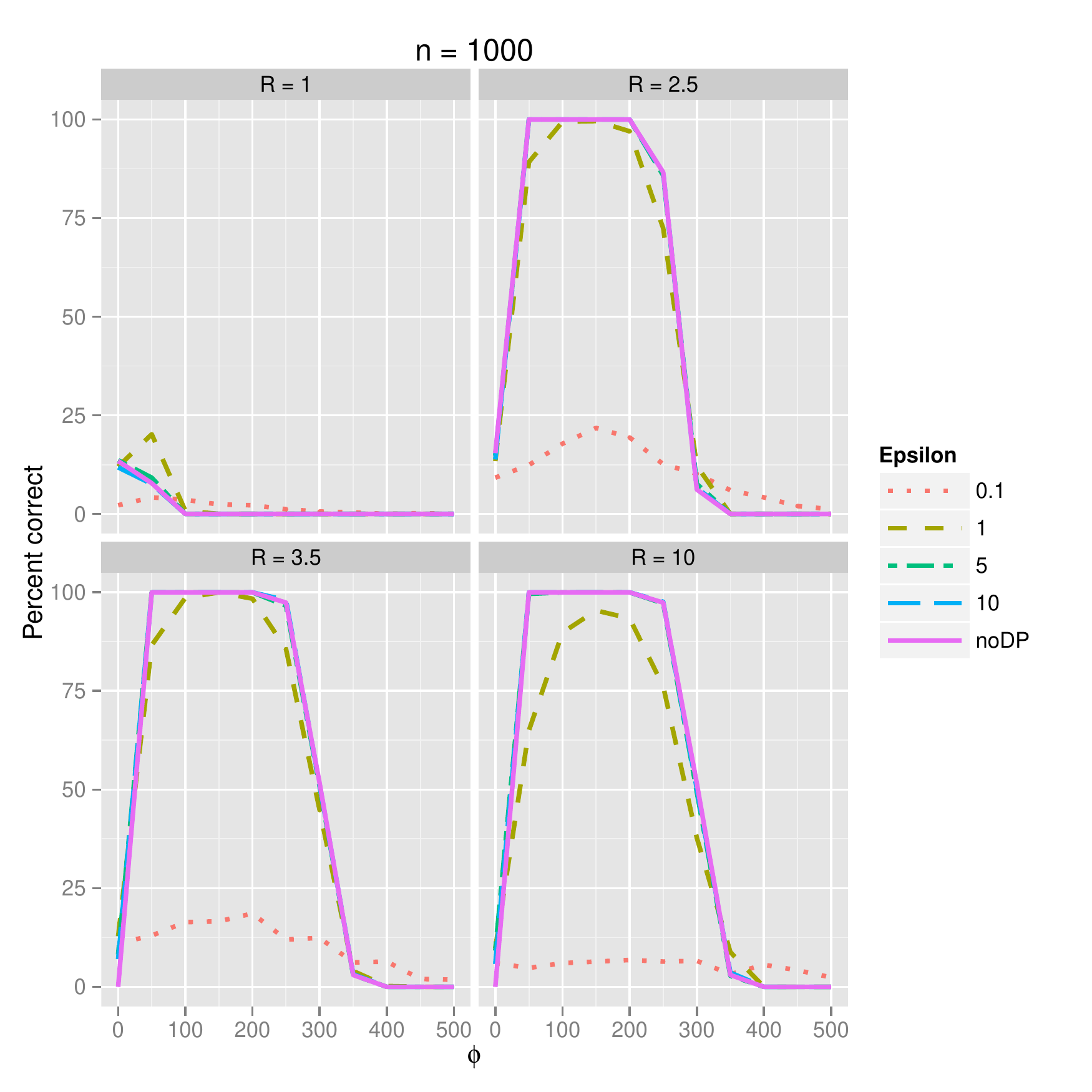}
  \caption{}
  \label{fig:sub2}
\end{subfigure}
\caption{Proportion of correct model selection from $500$ independent replications of Algorithm 1 selecting among the 63 possible models when the true model has $\beta = (1,1,1,0,0,0)$. Various values of $n$, $R$, $\phi$ and $\epsilon$ are illustrated, and $r$ is set to the maximum value of $Y$. There is a large range of $\phi$ for which the private procedure does as well as without privacy, and in fact picks the correct model. As $n$ increases, the task becomes easier. Note that the proper value of $\phi$ increases with $n$ as well. }
\label{fig:test}
\end{figure}

\begin{figure}[t]
\centering
\begin{subfigure}{.52\textwidth}
  \centering
  \includegraphics[scale = 0.4]{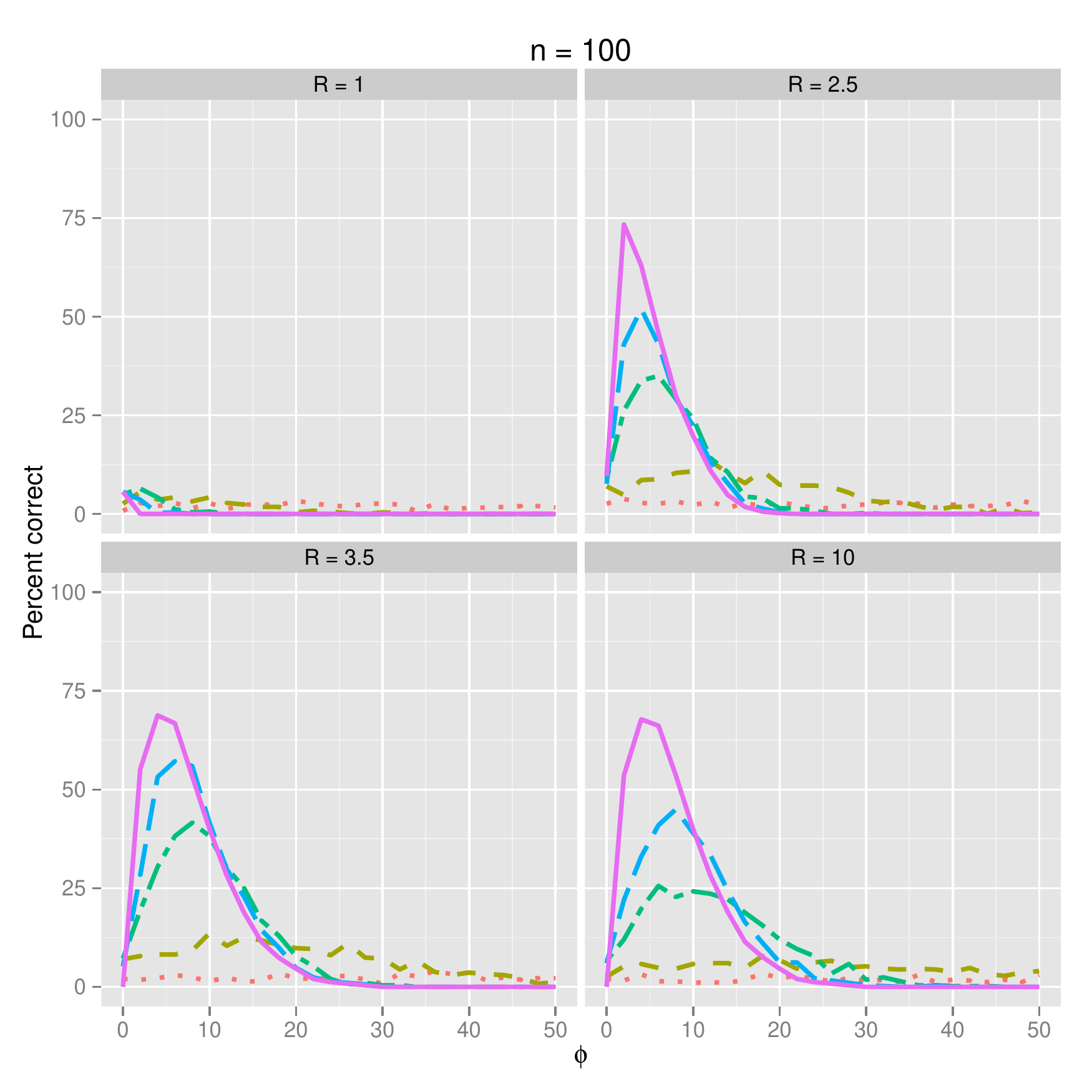}
  \caption{}
  \label{fig:sub1}
\end{subfigure}%
\begin{subfigure}{.55\textwidth}
  \centering
  \includegraphics[scale = 0.4]{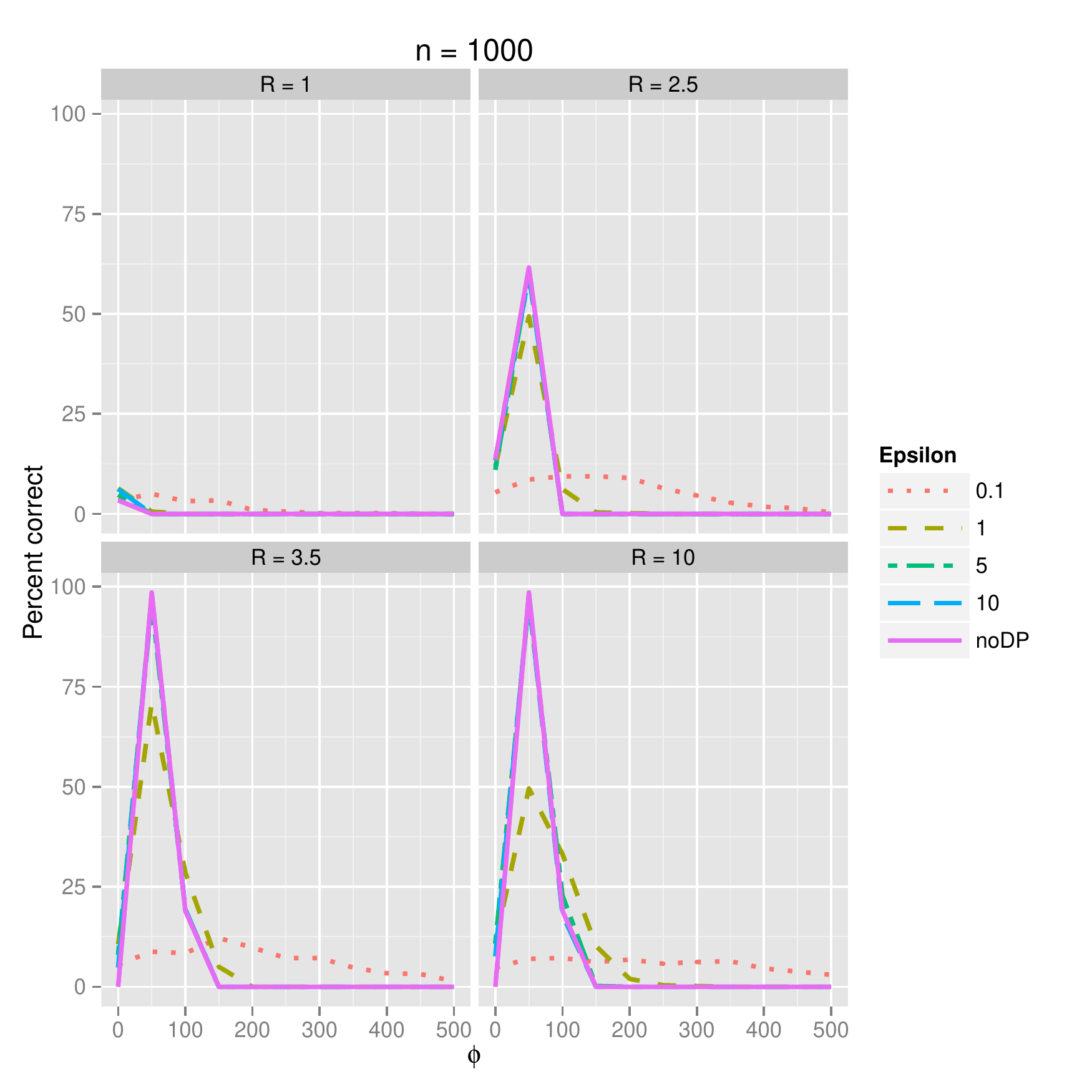}
  \caption{}
  \label{fig:sub2}
\end{subfigure}
\caption{Proportion of correct model selection from $500$ independent replications of Algorithm 1 selecting among the 63 possible models when the true model has $\beta = (1.5,1,0.5,0,0,0)$. Various values of $n$, $R$, $\phi$ and $\epsilon$ are illustrated, and $r$ is set to the maximum value of $Y$. The utility of the procedure depends crucially on the choices for $R$ and $\phi$. As $n$ increases, the task becomes easier. Note that the proper value of $\phi$ increases with $n$ as well. }
\label{fig:test}
\end{figure}
 \subsection{Application to Real Data Sets}
We first illustrate the results of model selection via penalized constrained least squares on a small data set of $97$ observations, then on a much larger one with hundreds of thousands observations. 

\paragraph{Prostate data set}
The prostate data set contains several clinical measures for $97$ men with prostate cancer. In this paper, our goal is to predict the level of a prostate specific antigen using five continuous variables: the volume of the cancer, the weight of the prostate, the age of the patient, the capsular penetration and the benign prostatic hyperplasia amount. Except for age, all variables are taken on the $log$ scale. We also rescale all of the variables to take values between $-1$ and $1$, which could be done in a differentially private way. 

We consider all possible main effects models for the model selection procedure, for a total of $63$ models to choose from. Model selection based on the penalized maximum likelihood,  without the constraint of differential privacy, selects the model with only two variables in addition to the intercept: the volume of the cancer, and the weight of the prostate. Model selection is then performed using Algorithm 1 for various choices of $R$, $\phi$ and $\epsilon$. We set $R=5.68$, obtained with the non-private algorithm, and we set $r$ to the maximum value of $Y$. 

\begin{figure}
\centering
\includegraphics[scale=0.8]{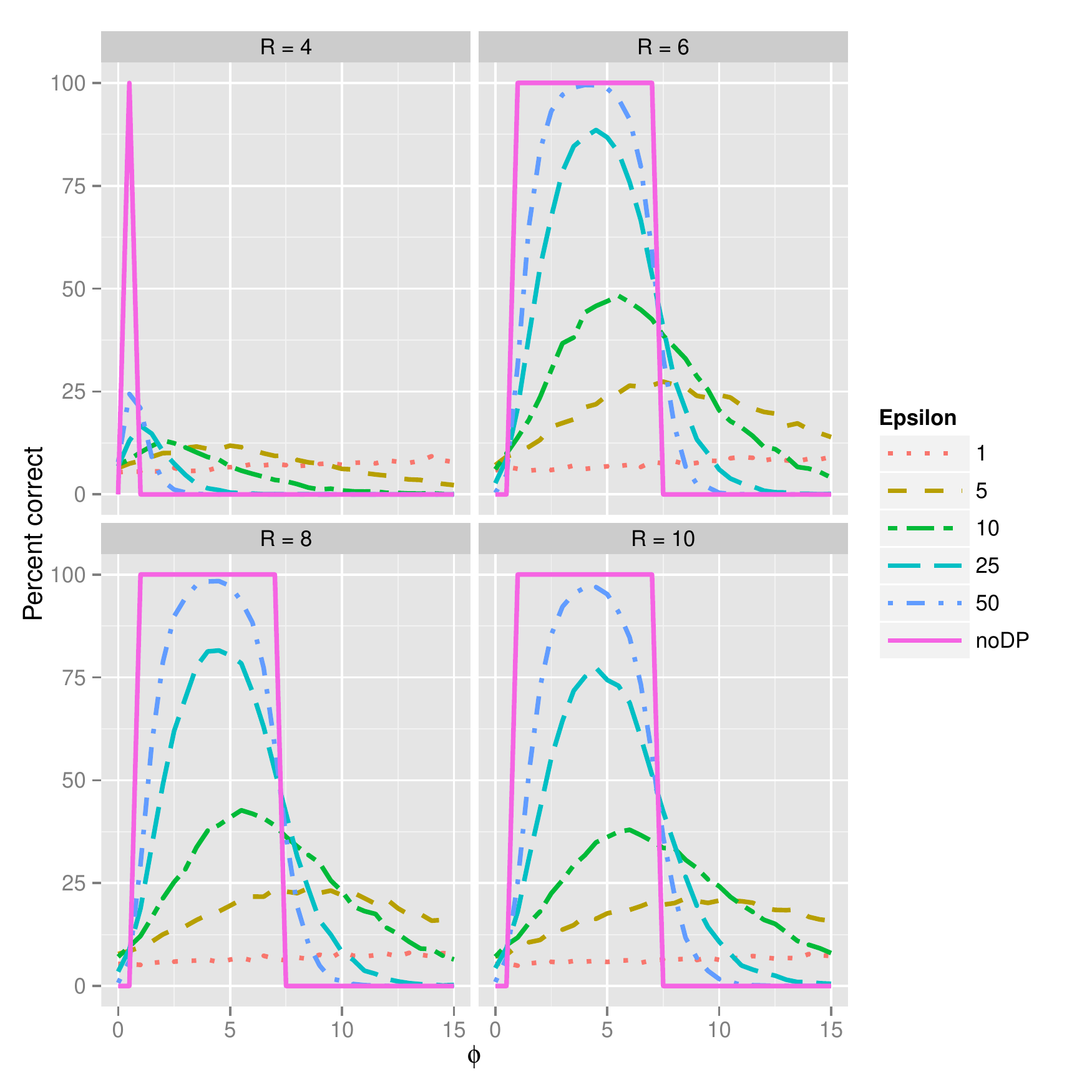}
\caption{Model selection on the prostate data set for the private and non-private procedures. Results show the proportion of times that the correct model is selected for $5000$ independent replications. Various values of $R$, $\phi$ and $\epsilon$ are illustrated. Because of the small sample size, the choice of $R$ and $\phi$ as more impact, and the utility is quite small for usual choices for $\epsilon$. }
\end{figure}

Figure 3 shows the proportion of times that the private and non-private procedures identify the correct model, for $5000$ independent replications. With such a small data set, the choice of $R$ and $\phi$ are crucial: even without differential privacy, using too large a value of $\phi$ does not identify the correct model. Due to the small sample size, the utility of the private procedure also decreases quite rapidly with decreasing $\epsilon$ which should offer more privacy. 


\paragraph{Housing dataset} 
Since the constrained optimization can be implemented based only on sufficient statistics, the private procedure scales very well for much larger data sets. The housing data set contains several variables measured on $348,189$ houses sold in San Francisco Bay Area between 2003 and 2006. In addition to the price of the sale, the data include the year of the transaction (an ordinal variable with 4 levels), the latitude and longitude of the house, the county in which it is located (a categorical variable with 9 levels), and a continuous measure of its size. We preprocess the data to remove houses with price outside the range $105$ to $905$ thousand dollars, and size larger than $3000$ sqft. We also combine some of the small counties into a new indicator variable. All predictors are also scaled so that they take values in $[-1,1]$. The resulting data set contains $235,760$ with $13$ variables. 

\begin{figure}
\centering
\includegraphics[scale=0.8]{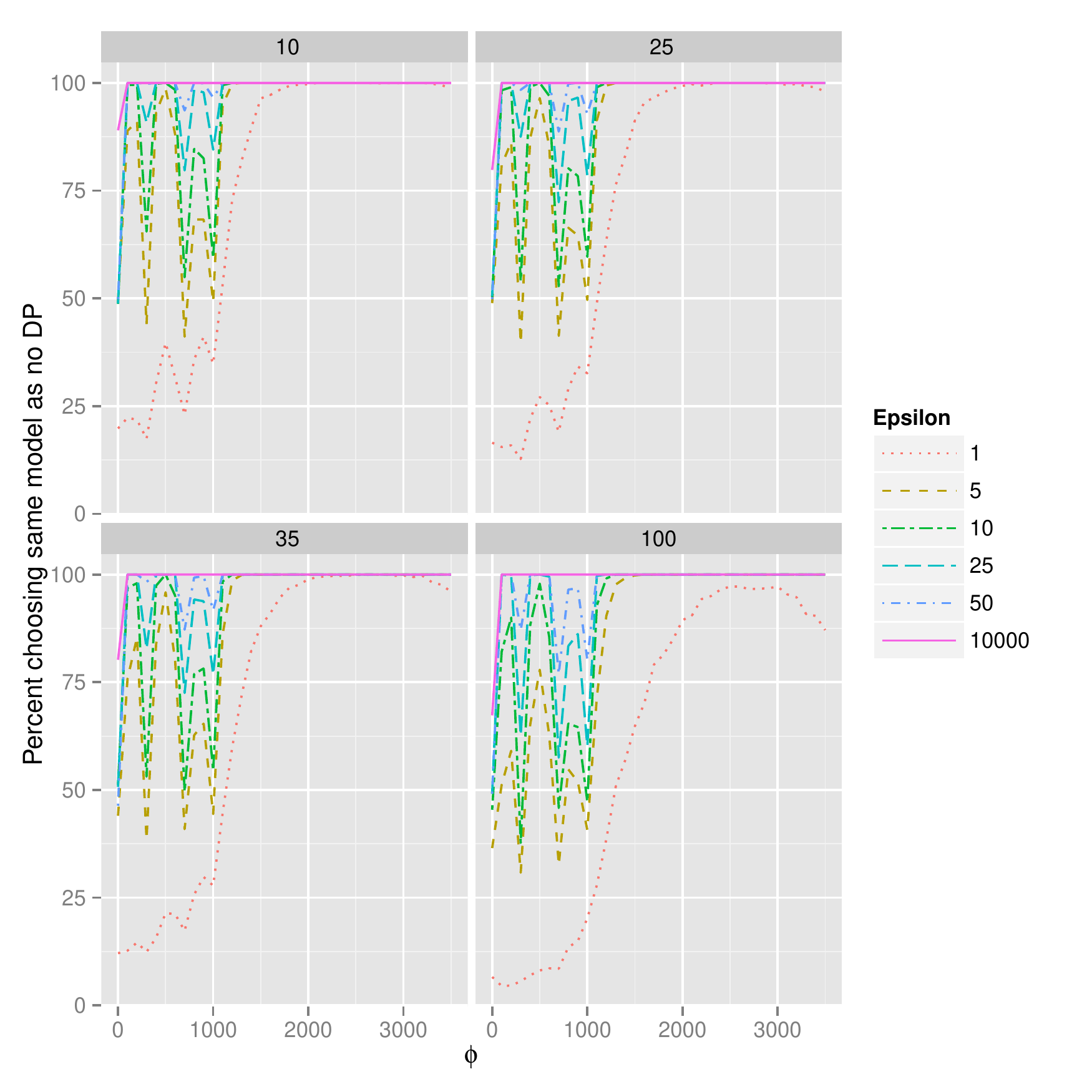}
\caption{Model selection on the housing data set. The y axis shows the proportion of time that the DP procedure selects the same model as the non-private version. Results represent $1000$ independent replications of Algorithm 1. Various values of $R$, $\phi$ and $\epsilon$ are illustrated. Even with small values of $\epsilon$, the private procedure returns the same model as the non-private one, for a large range of $\phi$. }
\end{figure}

We consider again models with main effects only, for a total $8191$ models to choose the best model from. Since the optimal model is not as clear as with the previous example, we do not directly compare the private procedure with a gold standard, but rather with its non-private counterpart. An algorithm which offers differential privacy but recovers the results of a non-private algorithm is successful. As above, we apply the model selection procedures with various values of $R$, $\phi$ and $\epsilon$, and use the maximum value of $Y$ as $r$. 

Figure 4 shows the agreement between the private and non-private procedure for various tuning parameters. With this large data set, the results are very positive: even for $\epsilon = 1$ we recover the same model with the differentially-private procedure as without the privacy requirement for $\phi$ chosen large enough. Note also that, as expected, the scale of $\phi$ is much larger than for the smaller {\em Housing} example. 


\section{Discussion}\label{sec:discussion}

With modern data acquisition and storage techniques allowing the collection and analysis of  huge amounts of personal information in a multitude of formats, protecting individual privacy inevitably becomes of crucial concern for modern data analysis. Although differential privacy offers mathematically strong and elegant privacy guarantee, the nature of such a conservative constraint in the context of everyday statistical analysis remains unclear. Previous works on statistical analysis with differential privacy mainly focused on simple statistical queries, such as location and scale statistics, regression coefficients, and simple hypothesis testing, and network analysis.  In this paper we considered the more challenging problem of model selection in the classical setting.  We showed that standard techniques for differentially private data analysis can be combined with known statistical tools such as penalized least square or information criteria to construct privacy-preserving model selection procedures with strong utility. We proposed two algorithms for this task, and proved privacy and utility results for each of them.  

Our procedures feature a double-regularization, as they include both  constrained estimation in the fitting step, and  penalization in the model comparison step. Thus the method involves two tuning parameters.  A key observation, illustrated in section 5, is that although we have proven good large-sample properties for a wide range of penalty parameter $\phi_n$ and $R$, the practical performance is sensitive to the particular choices for these parameters. In other words, these tuning parameters play a very important and unique role in designing differentially private procedure with good practical performance.  While in low-dimensional settings, the regularization parameter is not needed for statistical inference without privacy constraints, when privacy is a concern, a carefully chosen amount of regularization can lead to stable, low-sensitivity estimators even in the worst case. Appropriate choice of both tuning parameters thus reflects the need for some additional information in the data that will be useful for differentially private procedures, but not necessary in traditional inference methods.  It will be an interesting future topic to give a general characterization of such privacy related quantities, and to develop differentially private methods to estimate these quantities.

\section{Appendix: Proof details}\label{sec:proof-detail}
\subsection{Proofs}\label{sec:proof}
We first provide proofs of the theorems.
\paragraph{Proof of Proposition \ref{pro:generic}}
\begin{proof}
 \begin{align*}
   &P_{\omega}[\mathcal T(D,G(D))\in A] \\
   \le&P_\omega[\mathcal T(D,G(D))\in A, G(D)  \ge G^*(D) ] + \pr[G(D)< G^*(D)]\\
   \le&\int_{u\ge G^*(D)}P_\omega[\mathcal T(D,u)\in A] dP_{G(D)}(u)+\delta\\
   \le&\int_{u\ge G^*(D)}
    e^{\epsilon_2}P_\omega[\mathcal T(D',u)\in A] dP_{G(D)}(u)+\delta\\
   \le&\int_{u\ge G^*(D)}
    e^{\epsilon_2}P_\omega[\mathcal T(D',u)\in A] e^{\epsilon_1}dP_{G(D')}(u)+\delta\\
   \le & e^{\epsilon}P_\omega[\mathcal T(D',G(D'))\in A]+\delta. 
 \end{align*} 
\end{proof}

\paragraph{Proof of Lemma \ref{lem:local-sensitivity-RML}}
\begin{proof}
  Given a candidate model $M$,  the log likelihood is, ignoring
  constant terms,
  $$-2\ell^*(M;D)=n\log \left(
  n^{-1}\sum_{i=1}^n (Y_i-X_i^T\hat\beta_{M})^2\right).$$
  
  Let $\hat \beta'_{R,M}$ be the constrained least square estimator with input data set $D'$. 
  By boundedness of $Y_i$, $X_i$ and $\hat\beta_{R,M}$ we have
  $(Y_i-X_i^T \beta)^2\le (r+R)^2$ for all $\beta$ such that $\|\beta\|_1\le R$.
  Then using the fact that $\log(x/y)\le (x-y)/y$ for $x\ge y$, we have
  \begin{align*}
   &\left|2\ell^*_{R}(M;D')-2\ell^*_{R}(M;D)\right|
   =n\left|\log\left(\frac{\ell_R(M;D)}{\ell_R(M;D')}\right)\right|\\
    \le& \max\left\{
    \frac{(Y_n'-X_n'^T \hat\beta_{R,M})^2}{n^{-1}\sum_{i=1}^n(Y_i-X_i^T\hat\beta_{R,M})^2},
    \frac{(Y_n-X_n^T \hat\beta_{R,M}')^2}{n^{-1}\sum_{i=1}^n(Y_i'-X_i'^T\hat\beta_{R,M}')^2}
    \right\}\\
    \le &
    \frac{n(r+R)^2}{-2\ell_{R}(M;D)-(r+R)^2}\,. 
  \end{align*}
\end{proof}

Next we prove our main utility theorems.  The proofs of \Cref{thm:utility-pcls} and \Cref{thm:utility-local-sensitivity} rely on the following result, which is a consequence of simple linear algebra.
\begin{proposition}\label{pro:R_bound}
  We have
  $\|\hat\beta_{M}\|_1\le \sqrt{\bar d/ \kappa_0}r$ for all $M\in\mathcal M$. As a result,
  if $R\ge \sqrt{\bar d/ \kappa_0}r$ then $\hat\beta_{R,M}=\hat\beta_{M}$, 
  $\ell(M;D)=\ell_{R}(M;D)$, and $\ell^*_R(M;D)=\ell^*(M;D)$ for all $M\in\mathcal M$.
\end{proposition}
Proposition \ref{pro:R_bound} allows us to remove the $\ell_1$ constraint in our analysis. Although Proposition \ref{pro:R_bound} implies that
the $\ell_1$ constraint is inactive, such a constraint cannot be removed from the algorithms because it is used to bound the worst-case sensitivity of the estimators.

\paragraph{Proof of \Cref{thm:utility-pcls}}
\begin{proof}
  We prove for noisy minimization.  The argument can be adapted simply to cover the exponential mechanism.
  
  Define
  $$\Delta(M)=\left[-2\ell(M;D)+2\ell(M_0;D)\right]\,.$$
  Then according to Proposition \ref{pro:R_bound}, we can work directly with the unconstrained version.
  \begin{align*}
    P_{D,\omega}(\hat M \neq M)\le &
    \sum_{M'\in\mathcal M}P_{D,\omega}\left(-2\ell(M';D)+\phi_n|M'|+\frac{2(R+r)^2}{\epsilon}Z_{M'}\le\right.\\
    &\qquad\left. -2\ell(M_0;D)+\phi_n|M_0|+\frac{2(R+r)^2}{\epsilon}Z_{M_0}\right)\\
    \le & n^\alpha P_{D,\omega}\left\{
  \frac{2(R+r)^2}{\epsilon}(Z_1 - Z_2)\ge\sup_{M'\neq M_0}\left[\Delta_{M'}+(|M'|-|M_0|)\phi_n\right]
    \right\}\,.
  \end{align*}
  
For $M'\supset M_0$, using Claim 2 in \Cref{sec:proof-detail} we have
$\sup_{M':M'\supset M_0}\Delta_{M'}\ge -A\sigma^2(|M'|-|M_0|)\log n$ with
probability at least $1-\frac{\bar d-d_0}{\sqrt{2\pi A\log n}} n^{-c}$.
Thus with same probability we have
\begin{equation}\label{eq:utility-pcml-1}
  \sup_{M'\supset M_0}\Delta_{M'}+(|M'|-|M_0|)\phi_n
  \ge (|M'|-|M_0|)(\phi_n-A\sigma^2\log n)\ge \phi_n/2\,.
\end{equation}
For $M'\nsubseteq M_0$, we have by Claim 3 in \Cref{sec:proof-detail}, with
probability at least $1-\frac{1+\bar d}{\sqrt{2\pi A \log n}}n^{-c}$,
\begin{align}
  \sup_{M'\nsubseteq M} \Delta_{M'}+(|M'|-|M_0|)\phi_n
  \ge \frac{1}{2}\kappa_0 b_0^2 \sigma^2 n - |M_0|\phi_n
  \ge \phi_n/2\,.
\end{align}

Thus conditioning on $E_2^c\cap E_3^c$, which has probability at least $1-\frac{1+2\bar d}{\sqrt{2\pi A\log n}}n^{-c}$, we have
\begin{align*}
P_{\omega}(\hat M\neq M_0)&\le n^\alpha \exp\left(-\frac{\phi_n}{2}\frac{\epsilon}{2(R+r)^2}\right)\,.
\end{align*}

\end{proof}

\paragraph{Proof of Theorem \ref{thm:utility-local-sensitivity}}
\begin{proof}
Consider events $E_1$--$E_5$ as defined in \Cref{sec:proof-detail}.
We focus on the event $\left(\bigcup_{k=0}^5 E_k\right)^c$, which has
probability at least $1-3n^{-A/2}-2\bar d n^{-c}$.

Because $R\ge r\sqrt{\frac{\bar d}{\kappa_0}}\,$ by Proposition \ref{pro:R_bound} we have $-2\ell^*(M;D)=-2\ell^*_{R}(M;D)$ for all $M$ under consideration.

Next we bound the difference between $-2\ell^*(M;D)$ and $-2\ell^*(M_0;D)$.  Denote
$$
\Delta^*(M)=-2\ell^*(M;D)+2\ell^*(M_0;D)\,.
$$

In the case of $M\supset M_0$,
applying the fact that $-\log (1-x)\le x/(1-x)$ for $x\in (0,1)$ to
$$x=1-\frac{\sum_{i=1}^n(Y_i-X_i^T\hat\beta_M)^2}
{\sum_{i=1}^n(Y_i-X_i^T\hat\beta_0)^2}$$
we have
\begin{align}
0\le &-\Delta^*(M)=n\log\left(\frac{\sum_{i=1}^n (Y_i-X_i\hat \beta_0)^2}
{\sum_{i=1}^n (Y_i-X_i\hat \beta_M)^2}\right)
\le n \frac{-\Delta(M)}{-2\ell(M;D)}
\nonumber\\
\le &2A(|M|-|M_0|)\log n\,,\label{eq:Delta*_case1}
\end{align}
where the last step follows from the fact that we are not in the event $E_0\cup E_1\cup E_2$ defined in \Cref{sec:proof-detail}.

 In the case of $M\nsubseteq M_0$, 
\begin{align}
  \Delta^*(M)=&n\log\left(1+\frac{\Delta(M)}{-2\ell(M_0;D)}\right)
  \ge n\log\left(1+\frac{\kappa_0b_0^2}{4\sigma^2}\right)\nonumber
\end{align}
because we are in $E_0^c$ (which implies $-2\ell(M_0;D)\le 2n\sigma^2$ by Claim 1) 
and $E_3^c$
(which implies $\Delta(M)\ge n\kappa_0b_0^2\sigma^2/2$ by Claim 3).


Therefore, when
$$4A\log n \le \phi_n\le \frac{2}{3|M_0|}n\log\left(1+\frac{\kappa_0b_0^2}{4\sigma^2}\right)\,,
$$
we have
$$L_{R}^*(M;D)-L_{R}^*(M_0;D)\ge \phi_n/2,~~
\forall M\neq M_0\,.$$

On the event considered, we also have $G(D)\le 4(R+r)^2\sigma^{-2}$ by Claim 5.
Then we can bound the error probability by
\begin{align*}
  P_\omega[\hat M \neq M_0|D,G(D)]=&\sum_{M\in\mathcal M, M\neq M_0}P_\omega[\hat M=M]\\
  \le &  \sum_{M\in\mathcal M, M\neq M_0}\exp\left(-\frac{\epsilon}{4G(D)}(L_{R}^*(M;D)-L_{R}^*(M_0;D))\right)\\
  \le & n^{\alpha} \exp\left(-\frac{\epsilon \phi_n\sigma^2}{64(R+r)^2}\right)\,.
\end{align*}
 \end{proof} 

\subsection{Further proof details}\label{sec:detail}
Here we give details and summarize the multiple ``with high probability'' statements in the utility analysis.
Recall that
$$\Delta(M)=\left[-2\ell(M;D)+2\ell(M_0;D)\right]\,.$$

Let $A=2(\alpha+c)$, assume
\begin{align}
  \frac{n}{\log n}\ge&\max\left\{ \frac{64A\sigma^2}{\kappa_0b_0^2}, \frac{4\bar d\sigma^2}{\kappa_0b_0^2}, 8A\bar d \right\}\label{eq:detail-assumption-1}\\
  d_0\le &n / 8 \label{eq:detail-assumption-2}\\
  \epsilon \ge & \frac{(R+r)^2\left(\frac{A}{2}\log n +\log\frac{1}{2\delta}\right)}{\frac{n\sigma^2}{4}-(R+r)^2}
  \label{eq:detail-assumption-3}
\end{align}
We define the following events and give the corresponding probability bounds.
\begin{enumerate}
  \item Define
  \begin{align}
  \label{eq:event-E0}
  E_0 &\coloneqq \left\{D:-2\ell(M_0;D)\sigma^{-2}\ge (n-d_0)+\sqrt{2(n-d_0)A\log n}+A\log n\right\}\\
    \label{eq:event-l(M0)}
    E_1&\coloneqq\left\{D: -2\ell(M_0;D)\sigma^{-2}\le (n-d_0)-\sqrt{2(n-d_0)A\log n}\right\}\,.
  \end{align}
  \textbf{Claim 1.}
  $
  P_D(E_0)\le n^{-A/2},~~~P_D(E_1)\le n^{-A/2}.
  $
  Also using Eq. (\ref{eq:detail-assumption-1}), we have
  $-2\ell(M;D)\le 2n\sigma^2$ on $E_0^c$.
  \begin{proof}
    Without loss of generality, assume that $\sigma^2=1$.  Then
    $-2\ell(M_0;D)=\|\mathbf Y - P_{\Pi_0} \mathbf Y\|_2^2 = \|P_{\Pi_0^\bot}\mathbf W\|_2^2$ is a $\chi^2$ random variable
    with $n-d_0$ degrees of freedom.  The first claim follows from Lemma 1 of \cite{LaurentM00}. The second claim can be verified directly.
  \end{proof}

\item Define
\begin{equation}
  \label{eq:event-l(M)-large-set-upperbound}
  E_2\coloneqq\left\{\inf_{M\in\mathcal M, M\supset M_0} \frac{\Delta(M)}{\sigma^{2}(|M|-|M_0|)} < - A \log n\right\}
\end{equation}
\textbf{Claim 2.}
$
P_D(E_2)\le \frac{\bar d - d_0}{\sqrt{2\pi A\log n}}n^{-c}\,.
$
\begin{proof}
Because $M\supset M_0$,
  \begin{align*}
    0&\le -\Delta(M)= \|\mathbf Y - \mathbf X \hat\beta_{M_0}\|_2^2-\|\mathbf Y-\mathbf X\hat\beta_M\|_2^2\\
    &=\|\mathbf X\beta_0+\mathbf W - P_{\Pi_0}(\mathbf X \beta_0+\mathbf W)\|_2^2
    -\|\mathbf X\beta_0+\mathbf W - P_{\Pi_M}(\mathbf X\beta_0+\mathbf W)\|_2^2\\
    &=\|\mathbf W- P_{\Pi_0}(\mathbf W)\|_2^2
    -\|\mathbf W - P_{\Pi_M}(\mathbf W)\|_2^2= \|(P_{\Pi_M}-P_{\Pi_0})(\mathbf W)\|_2^2.
  \end{align*}
Because $M_0\subset M$, $P_{\Pi_M}-P_{\Pi_0}$ is a projection operator of dimension $|M|-|M_0|$.  Using a tail probability bound for Gaussian random variables, we  have, for all $A=2(\alpha+c)>0$,
  \begin{equation*}
    \pr\big[\Delta(M)\sigma^{-2}\le -A(|M|-|M_0|)\log n\big]\le \frac{(|M|-|M_0|)}{\sqrt{2\pi A\log n}}n^{-A/2}\le \frac{\bar d-d_0}{\sqrt{2\pi A\log n}}n^{-A/2}.\end{equation*}
    The desired result follows from union bound.
\end{proof}
\item For $M\nsupseteq M_0$, let $M_1= M_0\backslash M$, $M_2 = M_0\cap M$, $J_M^*=\sqrt{n\kappa_0 b_0^2|M_1|}$.  Define
\begin{align}
  E_3\coloneqq&\Big\{\inf_{M\in\mathcal M, M\nsupseteq M_0}
  \Delta(M) \le (J_M^*)^2-2\sigma J_M^*\sqrt{A\log n}-(|M|-|M_2|)
  A\log n
  \Big\}\,.\label{eq:event-l(M)-smaller-set-lower-bound}
\end{align}
\textbf{Claim 3.}
$$
P_D(E_3)\le \frac{1+\bar d}{\sqrt{2\pi A\log n}}n^{-c}\,.
$$
and under $E_3^c$ and equation (\ref{eq:detail-assumption-1})
$$
   \Delta(M) \ge \frac{1}{2}\kappa_0b_0^2 \sigma^2 n\,,~~\forall M\in\mathcal M,~ M\nsupseteq M_0\,.
$$
\begin{proof}
    Denote
$\beta_{0,M}$ the vector that agrees with $\beta_0$ on $M$ and 0 elsewhere.
\begin{align*}
  &\Delta(M)=\|\mathbf Y-P_{\Pi_M}(\mathbf Y)\|_2^2 - \|P_{\Pi_0^\bot}(\mathbf{W})\|_2^2\\
  =&\|P_{\Pi_M^\bot}(\mathbf X\beta_{0})+P_{\Pi_M^\bot}(\mathbf{W})\|_2^2-\|P_{\Pi_0^\bot}(\mathbf{W})\|_2^2\\
  =&\|P_{\Pi_M^\bot}(\mathbf X_{M_1}\beta_{0,M_1})\|_2^2+2\langle P_{\Pi_M^\bot}
  (\mathbf X_{M_1}\beta_{0,M_1}), P_{\Pi_M^\bot}(\mathbf{W}) \rangle +\|P_{\Pi_M^\bot}(\mathbf{W})\|_2^2-
  \|P_{\Pi_0^\bot}(\mathbf{W})\|_2^2\\
  \ge&\|P_{\Pi_M^\bot}(\mathbf X_{M_1}\beta_{0,M_1})\|_2^2+2\langle P_{\Pi_M^\bot}
  (\mathbf X_{M_1}\beta_{0,M_1}), \mathbf{W} \rangle-
  \|P_{\Pi_{0,M} \cap \Pi_0^\bot}(\mathbf{W})\|_2^2\,,
\end{align*}
where $\Pi_{0,M}$ is the linear subspace spanned by $X_{M_0\cup M}$.

Let $J = \|P_{\Pi_M^\bot}(\mathbf X_{M_1}\beta_{0,M_1})\|_2$.  Then
the second term in the above equation is
distributed as $N(0, \sigma^2 J^2)$.
By the eigenvalue condition, we have
$J^2 \ge \kappa_{0} n b_0^2 |M_1|$, where
$\kappa_{0}$ is the minimum sparse eigenvalue
and $b_0$ is a lower bound of the signal level. To see this,
observe that $M\cap M_1=\emptyset$ and
\begin{align*}
J=\left\|
P_{\Pi_M^\bot}(\mathbf X_{M_1}\beta_{0,M_1})
\right\|
=&
\left\|
\left(\mathbf X_{M_1},\mathbf X_{M}\right)\left(
\begin{array}
  {c}
  \beta_{0,M_1}\\
  -(\mathbf X_M^T\mathbf X_M)^{-1}\mathbf X_M^T \mathbf X_{M_1}\beta_{0,M_1}
\end{array}
\right)
\right\|\\
\ge &
\sqrt{n\kappa_0}\left\|
\left(
\begin{array}
  {c}
  \beta_{M_1}\\
  -(X_M^TX_M)^{-1}X_M^T X_{M_1}\beta_{M_1}
\end{array}
\right)
\right\|\\
\ge &
\sqrt{n\kappa_0}\left\|
  \beta_{M_1}
\right\|\ge \sqrt{n\kappa_0|M_1|}b_0,
\end{align*}
where the first inequality uses the sparse eigenvalue condition.

Note that the dimension of $\Pi_M\cap \Pi_0^\bot$
is at most $|M|-|M_2|$.
Let $J^*\coloneqq \sqrt{\kappa_0 n b_0^2}$.  Using \eqref{eq:detail-assumption-1} we know that
$J\ge J^*\ge 4\sqrt{A\log n}$.
 Then with probability at least $1-(1+|M|-|M_2|)n^{-A/2}/\sqrt{2\pi A\log n}\le 1-\frac{1+\bar d}{\sqrt{2\pi A\log n}}n^{A/2}$, we have
\begin{align}
\label{eq:lower_bound_Delta_case2}
\Delta(M)\sigma^{-2}\ge
    &  J^2 - 2\sigma J\sqrt{A\log n} - (|M|-|M_2|)\sigma^2 A\log n\\
\ge &  (J^*)^2 - 2\sigma J^* \sqrt{A\log n} - \sigma^2\bar d A\log n\\
\ge & \frac{1}{2}\kappa_0 b_0^2 n
\end{align}
where the last inequality uses eq. (\ref{eq:detail-assumption-1}).  The claim follows from union bound.
\end{proof}
  \item Define
  \begin{equation}
    \label{eq:event-l(M)-lowerbound}
    E_4\coloneqq\left\{D: \inf_{M\in\mathcal M} -2\ell(M;D)\le \sigma^2 n / 2 \right\}\,.
  \end{equation}
\textbf{Claim 4.}  $E_4\subseteq E_1\cup E_2\cup E_3\,.$

To see this, first note that on
$(E_1\cup E_2\cup E_3)^c$, we have
$\min_{M\in\mathcal M}-2\ell(M;D)\ge \sigma^2(n-d_0-\sqrt{2(n-d_0)A\log n}-A(\bar d-d_0)\log n)\ge n\sigma^2/2$
using eq. (\ref{eq:detail-assumption-1})--(\ref{eq:detail-assumption-2}).

\item Define
\begin{equation}
  \label{eq:event_G(D)_bound}
  E_5\coloneqq\{z_G<-A\log n /2\}\,.
\end{equation}
\textbf{Claim 5.} $P_\omega(E_5)= \frac{1}{2}n^{-A/2}\le n^{-A/2}$.
On $E_4^c\cap E_5^c$, we have, using eq. (\ref{eq:detail-assumption-1})--(\ref{eq:detail-assumption-3}),
$$
G(D)\le \frac{4(R+r)^2}{\sigma^2}\,.
$$
The proof is elementary and omitted.
\end{enumerate}


\bibliographystyle{apa-good}
\bibliography{paper}
\end{document}